\documentclass[runningheads]{llncs}

\usepackage{xcolor}
\usepackage{amssymb}
\usepackage{amsmath}
\usepackage{graphicx}
\usepackage{epstopdf}
\usepackage{subfigure}

\usepackage{amsfonts}

\usepackage{siunitx}
\usepackage{tikz} % To generate the plot from csv
\usepackage{pgfplots}

\usepackage{makecell}

\pgfplotsset{compat=newest} % Allows to place the legend below plot
\usepgfplotslibrary{units} % Allows to enter the units nicely

\usepackage{url}
\urldef{\mailsa}\path|{alfred.hofmann, ursula.barth, ingrid.haas, frank.holzwarth,|
\urldef{\mailsb}\path|anna.kramer, leonie.kunz, christine.reiss, nicole.sator,|
\urldef{\mailsc}\path|erika.siebert-cole, peter.strasser, lncs}@springer.com|    
\newcommand{\keywords}[1]{\par\addvspace\baselineskip
\noindent\keywordname\enspace\ignorespaces#1}

\usepackage{multirow}
\usepackage{adjustbox}
\usepackage{verbatim}
\usepackage[ruled]{algorithm2e} %linesnumbered

\usepackage[noend]{algpseudocode}
\usepackage{mathtools}
\DeclarePairedDelimiter\floor{\lfloor}{\rfloor}

\begin{document}

\mainmatter  % start of an individual contribution

\title{ Resource Optimized Neural Architecture Search for 3D Medical Image Segmentation }
% for lung nodule detection

%\titlerunning{DeepLung : Semantic Lung Segmentation in HRCT using DCNN}


\author{Woong Bae\inst{1}, Seungho Lee\inst{1}, Yeha Lee, Beomhee Park, Minki Chung, and Kyu-Hwan Jung\inst{*}\\} %1{Bae, Woong}, 2{Lee, Seungho}, 3{Lee, Yeha}, 4{Park, Beomhee}, 5{Chung, Minki}, 6{Jung, Kyu-Hwan} 
\institute{VUNO Inc., Seoul, South Korea\\\{iorism, bleaf.lee, yeha.lee, beomheep, brekkanegg, khwan.jung\}@vuno.co}
%\email{{iorism, bleaf.lee, yeha.lee, beomheep, brekkanegg, \\khwan.jung}@vuno.co\\}
\footnotetext[1]{The two authors contributed equally to this paper.}


\toctitle{Lecture Notes in Computer Science}
\tocauthor{Authors' Instructions}
\maketitle

\begin{abstract}
Neural Architecture Search (NAS), a framework which automates the task of designing neural networks, has recently been actively studied in the field of deep learning.
However, there are only a few NAS methods suitable for 3D medical image segmentation. 
Medical 3D images are generally very large; thus it is difficult to apply previous NAS methods due to their GPU computational burden and long training time.
We propose the resource-optimized neural architecture search method which can be applied to 3D medical segmentation tasks in a short training time (1.39 days for 1GB dataset) using a small amount of computation power (one RTX 2080Ti, 10.8GB GPU memory).
Excellent performance can also be achieved without retraining(fine-tuning) which is essential in most NAS methods.
These advantages can be achieved by using a reinforcement learning-based controller with parameter sharing and focusing on the optimal search space configuration of macro search rather than micro search. 
Our experiments demonstrate that the proposed NAS method outperforms manually designed networks with state-of-the-art performance in 3D medical image segmentation.

\end{abstract}

%%%%%%%%%%%
\section{Introduction}\label{intro}
Research using deep neural networks for 3D medical image segmentation has exploded over the last few years, producing excellent methods such as U-Net \cite{ronneberger2015u} and deep supervision \cite{dou20163d}.
However, the performance of these methods is highly influenced by manual tasks such as post-processing, hyperparameter tuning, and designing an optimal architecture.
In particular, fine-tuning the hyperparameters and designing the best architecture require a great deal of time and computational power. 
To reduce these manual tasks, research in automated machine learning has been actively carried out for natural image processing tasks 
\cite{bender2018understanding,liu2018darts,pham2018efficient}.
Despite their success in natural image processing, these methods are difficult to apply to the segmentation of 3D medical image with high dimensions which requires enormous computational power.
Automated methods developed for 3D medical image segmentation do not yet match state-of-the-art performances achieved by manually designed methods
\cite{mortazi2018automatically,Sungwoong_SCNAS}.

In this paper we propose a resource optimized neural architecture search for 3D medical image segmentation (RONASMIS) which takes a short training time (1.39 days for 1GB dataset) and requires a small amount of GPU computational power (one RTX 2080Ti with 10.8GB).
The proposed framework differs from previous NAS frameworks by focusing on macro search rather than micro search, exploiting the characteristics of 3D medical images.
We avoid retraining the network from scratch after finding the optimal architecture by continuously training the child network during the architecture search process without re-initializing the child network's weights.
GPU memory is efficiently utilized by using addition-based skip connections and replacing depth-wise convolutions with normal convolutions.
The search space only includes elements that significantly impact the final performance of 3D medical image segmentation, reducing the architecture complexity and GPU memory usage. To the best of our knowledge, our framework is the first to outperform state-of-the-art results in 3D medical image segmentation.

\begin{figure} \centering %[h]
	\includegraphics[scale=0.37]{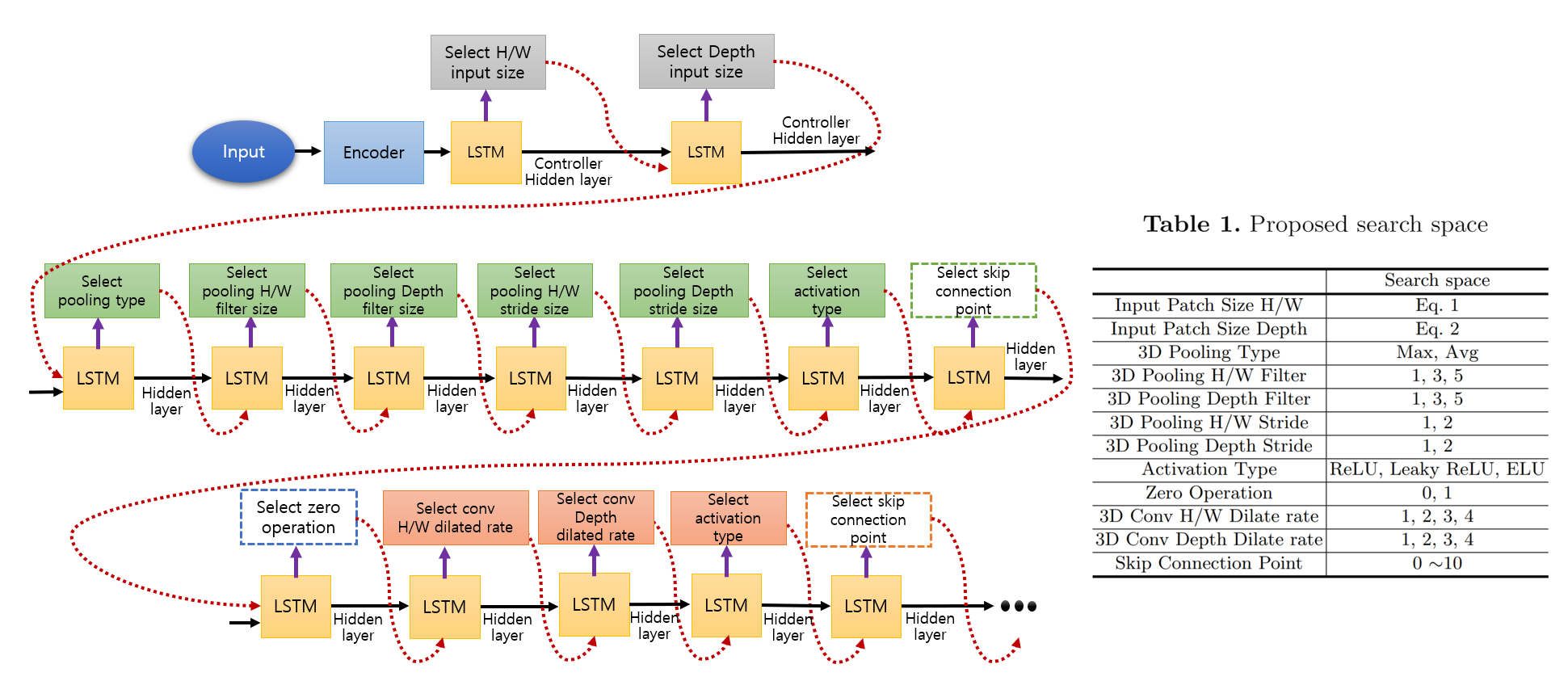}  %0.28
	\caption{The proposed RNN based controller and search space. The left figure shows the structure of the proposed controller with dotted boxes indicating that there is a point not applicable in certain sections. Table 1. shows the proposed search space.}
    \label{fig:controller_main}
\end{figure}

\section{Method}
\subsection{Resource-Optimized Search Space for Anisotropic 3D Medical Image and the Basic Architecture} 
\textbf{Proposed Search Space : }The search space we propose is shown in Table 1. of Figure \ref{fig:controller_main}. 
Our search space is constructed differently from other NAS papers based on four main reasons:
\begin{enumerate}
    \item We include input patch sizes and the amount of down sampling in the search space taking into consideration the fact that most 3D medical images have anisotropic shapes. Images in our training data have shapes $4 \times 155 \times 240 \times 240$, $1\times[90 \sim 130]\times320\times320$, and $2\times[11\sim24]\times[256\sim384]\times[256\sim384]$ in order of channel, depth, height, and width. The variation of input patch sizes and the number of down sampling operations were also considered by \cite{isensee2018nnu} which won the medical segmentation decathlon (MSD).
    \item To effectively acquire the receptive field and preserve features' spatial resolution information as much as possible in the encoding process, we include the amount of pooling and dilation rate of 3D convolution in the search space.
    \item One main goal of this paper is efficient utilization of GPU memory. A micro search space requires extensive GPU memory which is already burdensome for 3D medical image segmentation.
    We thus resort to focusing on a macro search space whereas other NAS methods consider either both micro and macro search space or just the prior
    \cite{bender2018understanding,liu2018darts,Sungwoong_SCNAS}.
    \item Some studies show that connections between global features are pivotal \cite{shah2018ms}. Most NAS research considering micro search spaces receive inputs from one or two adjacent cells. We instead include skip connection points in our search space to maximize the effect of skip connections across the network.
\end{enumerate}

\begin{comment}
%1
The first reason is that most 3D medical images have anisotropic shapes. 
Images in our training data have shapes $4 \times 155 \times 240 \times 240$, $1\times[90-130]\times320\times320$, and $2\times[11-24]\times[256-384]\times[256-384]$ in order of channel, depth, height, and width respectively. The variation of input patch sizes and the number of down sampling operations were also considered by \cite{isensee2018nnu} which won the medical segmentation decathlon (MSD).
%2
The second reason is to effectively acquire the receptive field and preserve features' spatial resolution information as much as possible in the encoding process. This can be achieved by using dilated convolution. We thus include the amount of pooling and dilation rate of 3D convolution in the search space.
%3
The third reason is to efficiently utilize GPU memory. A micro search space requires extensive GPU memory which is already cumbersome for 3D medical image segmentation. This motivates us to use a macro search space in contrast to other NAS methods \cite{bender2018understanding,liu2018darts,Sungwoong_SCNAS}.
The last reason is to maximize the effect of skip connections across the entire network. Most NAS papers focusing on micro search receive inputs from one or two adjacent cells. 
However, some studies show that connections between global features are pivotal\cite{shah2018ms}.
\end{comment}

Further, other NAS methods often apply skip connections to $1\times1$ convolution after concatenation with other inputs. 
This increases the number of parameters, resulting in more GPU memory usage than addition. As a remedy we use elementwise-sum based skip connection for macro search.

Most researchers manually determine the best 2D or 3D input patch size, type of down sampling operations along with their stride, and skip connection methods, taking into account available GPU memory and receptive field size \cite{dou20163d,drozdzal2016importance,isensee2018nnu}. Cascaded learning methods are often used as an alternative; however, the training time takes longer and still requires hyper-parameter tuning \cite{wang2017automatic}. Instead, we focus on constructing a resource optimized search space to reduce the time required to tune hyperparameters. 

Three activation functions and two pooling operation types are included in our search space because they do not use additional GPU memory and may affect the segmentation performance depending on the task. To prevent overfitting of the child network, we also add a drop-path regularization (zero operation) which disables some operations or skip-connections at a specific node in the search space. This also helps the controller reliably construct the architecture \cite{bender2018understanding}, \cite{pham2018efficient}. 

We replace depth-wise convolution used in most NAS frameworks with normal convolution because the latter utilizes GPU memory more efficiently. It is commonly believed that depth-wise convolution saves more memory because it uses less parameters and requires fewer arithmetic operations. However, there are implementation issues in open source frameworks such as PyTorch which cause depth-wise convolution to inefficiently use up cache memory and more GPU memory when performing backpropagation.

Eqs. \eqref{eq:SS_HW} and \eqref{eq:SS_D} show how to determine the input patch size in the search space given the the input training image dimensions. 
\begin{equation}\label{eq:SS_HW}
\text{Search} \, \text{Space} \, \text{of} \, \text{Patch Size H/W} =  \floor*{\frac{ \max(H, W)  }{S^4}} *{S^4} - {S^4} * \{0, 1, 2, 3, 4\}
\end{equation}
\begin{equation}\label{eq:SS_D} 
\text{Search} \, \text{Space} \, \text{of} \, \text{Patch Size Depth} =  \floor*{\frac{ D }{S^4}}  *{S^4} - {S^4} * \{0, 1, 2, 3, 4\}
\end{equation}
where $H, W, D$ are the median height, width, and depth sizes of training images respectively and $S$ is the stride parameter for each stage. 
The height and width patch sizes are set equal and the depth is determined independently. The search space cardinality of each element is limited to $5$ for any given input patch size. 

When an image dimension in the intermediate stages becomes smaller than the largest pooling size, the search space for pooling stride is modified to smaller values.
For example, the smallest depth size of prostate 3D images in the MSD challenge is $11$. When using this particular image, the search space for pooling stride in the depth direction is set to $\{1\}$. This is necessary because the range of image sizes is unknown prior to training.

\begin{figure} \centering %[h]
	\includegraphics[scale=0.37]{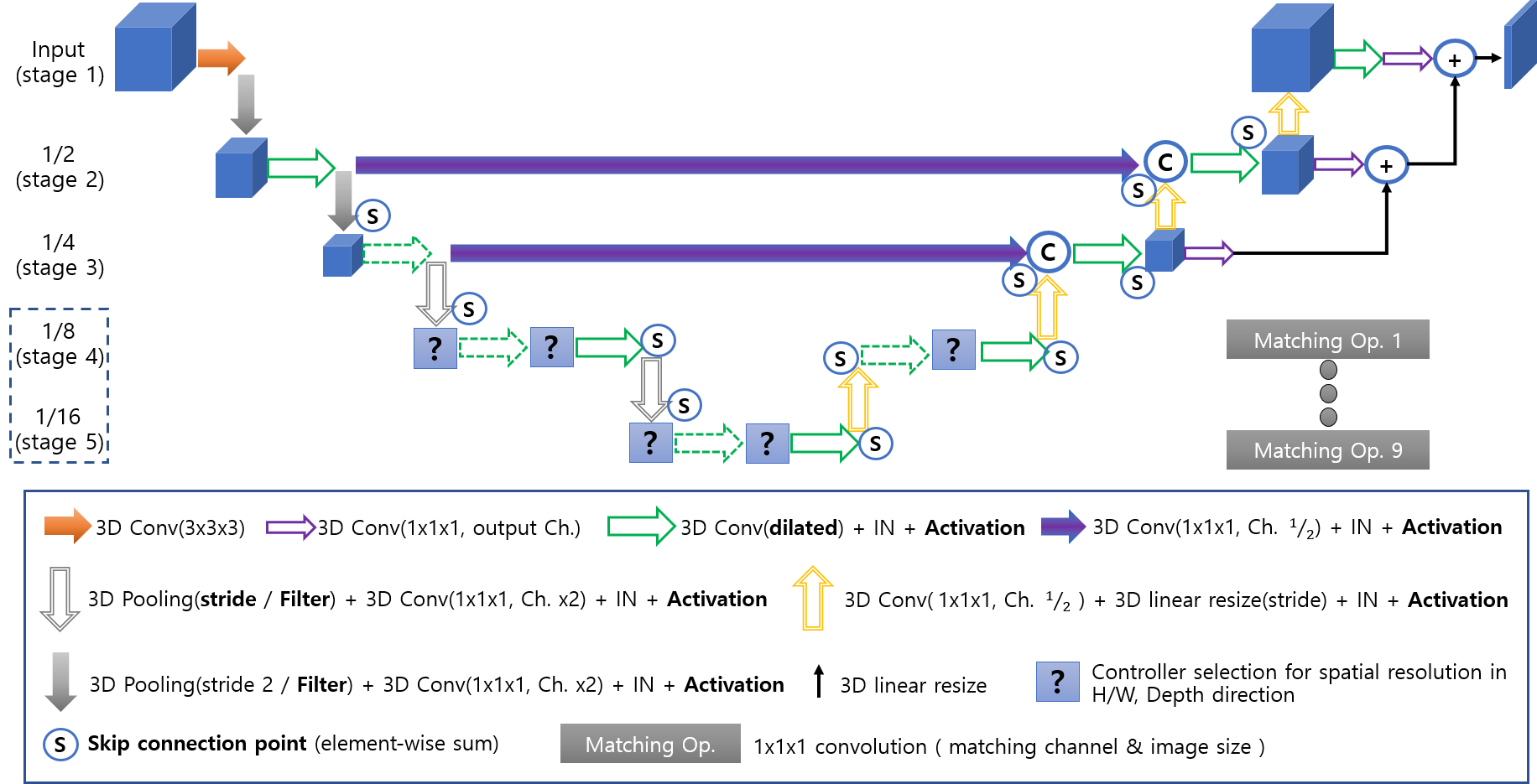}  %0.28
	\caption{The proposed base architecture.}
    \label{fig:base_network}
\end{figure}

\noindent\textbf{The Proposed Base Architecture :}
We use an architecture modified from U-Net\cite{ronneberger2015u} for our base architecture as shown in Figure \ref{fig:base_network}. The architecture combines the decoder $1\times1\times1$ convolution skip connections in DeepLabV3+\cite{chen2018encoder} and the deep supervision scheme proposed in \cite{dou20163d}. Batch normalization is replaced with instance normalization to account for GPU memory.

Filled purple arrows indicate $1\times1\times1$ convolution skip connections that control the amount of information transfer between the encoder and decoder. The number of channels is halved, and the output is concatenated with the decoder's output.
The unfilled purple and black arrows are a combination of 3D resize and element-wise-sum of outputs from stages 1, 2, and 3 which is a modified deep supervision method. 
The parameters considered in the search space are marked bold in the box below the architecture diagram. Question marks denote positions in which feature sizes are determined by the pooling stride parameters in stages 3 and 4.
The dotted green arrows indicate whether a zero operation should be used or not.
Skip connections which send and receive features are marked with S circles.

When skip-connected features differ in channel sizes, most NAS frameworks concatenate all skip-connected features and apply $1\times1\times1$ convolution to each layer.
On the other hand, we use a matching operation where $1\times1\times1$ convolutions are applied to all skip connected features with the corresponding channel size to decrease GPU memory usage and network parameters, and then perform an element-wise-sum. 
We expect this to transfer all import information contained in the features since features with various spatial resolutions are used during architecture search.

All 3D convolution kernels are $1 \times 1 \times 1$ or $3 \times 3 \times 3$ and the controller selects the convolution dilation rate. To stabilize training, the dilation rate of 3D convolution at stage 1 in the encoder part is fixed to 1, and the pooling strides at stage 1 and 2 are fixed to $2$. The features in the inner-most stage can take resolutions ranging from $1/16$ to $1/4$ of the input image's resolution.

\subsection{Training the Controller's Parameters for Architecture Selection}\label{sec:Method}
We use a parameter sharing based reinforcement learning proposed by ENAS\cite{pham2018efficient} to train the controller.
Parameter sharing is an excellent method which trains the controller by receiving a reward just from the newly constructed architecture without retraining the child network from scratch. This reduces the time necessary to reach the global optimum although performance may be affected by the initial architecture, number of episodes, number of training epochs for the child network, moving average baseline parameter, and the entropy's regularization coefficient. We assumed that the combination of sequential actions constitutes the optimal architecture. Thus, we distinguish distinct operations by adding unique values to the LSTM inputs.
On the other hand, the reason we do not use the recently proposed differentiable NAS scheme is that the method uploads all operations on the GPU, requiring many expensive GPUs.

At each episode the controller creates 20 child networks and observes the corresponding validation patient-wise dice scores of each network which are used as a reward to train the controller. The 3D dice score is calculated after thresholding pixel values lower than $0.5$ to $0$. The next generation architecture is then determined by the sequence of actions taken by the controller. The generated architecture is be trained using dice loss \cite{drozdzal2016importance}.

\section{Implementation Details and Experimental Results}
\subsection{Data and Implementation Details}\label{sec:results}
\textbf{Dataset : } 
We conducted experiments on the brain, heart, and prostate 3D medical images used in the medical segmentation decathlon challenge(MSD, \url{http://medicaldecathlon.com/}).
This dataset contains images with significantly varying 3D shape and channel sizes, making it appropriate for confirming our NAS method's performance.
Because the labels for the test dataset are not publicly available, we evaluate performance with 5-fold cross validation as in \cite{isensee2018nnu,Sungwoong_SCNAS}, not on the test dataset.

\noindent\textbf{Implementation Details : } 
We preprocess the data by applying Z-score normalization to each channel independently. For the brain data, we cropped the images based on their non-zero minimum value of each dimension because many voxels contain zeros.
The learning rate and weight decay of the child network were fixed as $0.001$ and $0.00001$ respectively without any learning rate decay methods. The controller also used a fixed learning rate and weight decay of $0.001$ and $0.000001$ respectively, with entropy regularization coefficient set as $0.0001$. The ADAM optimizer was used for both the child and controller networks. 
Considering training time, the controller's training epochs were $150$, $500$, and $500$ for brain, heart, and prostate tasks respectively.
The child network was trained for $3$ epochs per episode.
To receive a more meaningful reward in the first epoch, the first child network arbitrarily used the architecture with the maximum input patch size including all skip connections and operations. 

It is known that changes in batch normalization statistics hinder the training process early on when using drop-path regularization on NAS \cite{bender2018understanding}. However, our training remains unaffected because batch normalization is replaced with instance normalization.
The batch size was chosen based on available memory fixed as $2, 1$, and $4$ for brain, heart, and prostate tasks respectively. The code was implemented using PyTorch 1.0.0.

\subsection{Experimental Results} %\ref{tab:DICE_result}. 
Our comparison baselines include nnUnet\cite{isensee2018nnu} and SCNAS\cite{Sungwoong_SCNAS} as shown in Table 2. nnUnet achieves state-of-the-art performance and holds first place in the MSD competition. 
SCNAS is a gradient-based AutoML network with a micro search space developed for 3D medical image segmentation tasks.
Our method achieves performance exceeding that obtained by SCNAS and nnUnet. Furthermore, this is achieved without using much memory, being small enough to be trained on just one RTX 2080Ti (10.8GB GPU memory) and is computationally cheap, taking $3.1$, $1.39$, and $0.35$ days to train on the brain, heart, prostate tasks respectively.
 
In the MSD challenge, most teams employ additional methods such as ensembling multiple neural networks, abundant data augmentation techniques, Test Time Augmentation (T.T.A), and post-processing methods.
In contrast, we only use horizontal flip in the 2D axial dimension for data augmentation to evaluate the proposed NAS performance.
We also did not use use any pre-trained model weights, ensembling, T.T.A, nor post-processing methods. 
Furthermore, most teams use the 50\% overlapped patch-wise inference method to improve performance\cite{isensee2018nnu,Sungwoong_SCNAS}. 
This method increases the inference time and requires more computation. Instead, we use the one-shot inference method to process the whole image in a single forward pass for quick inference although this can degrade performance.

\begin{figure} \centering %[h]
	\includegraphics[scale=0.36]{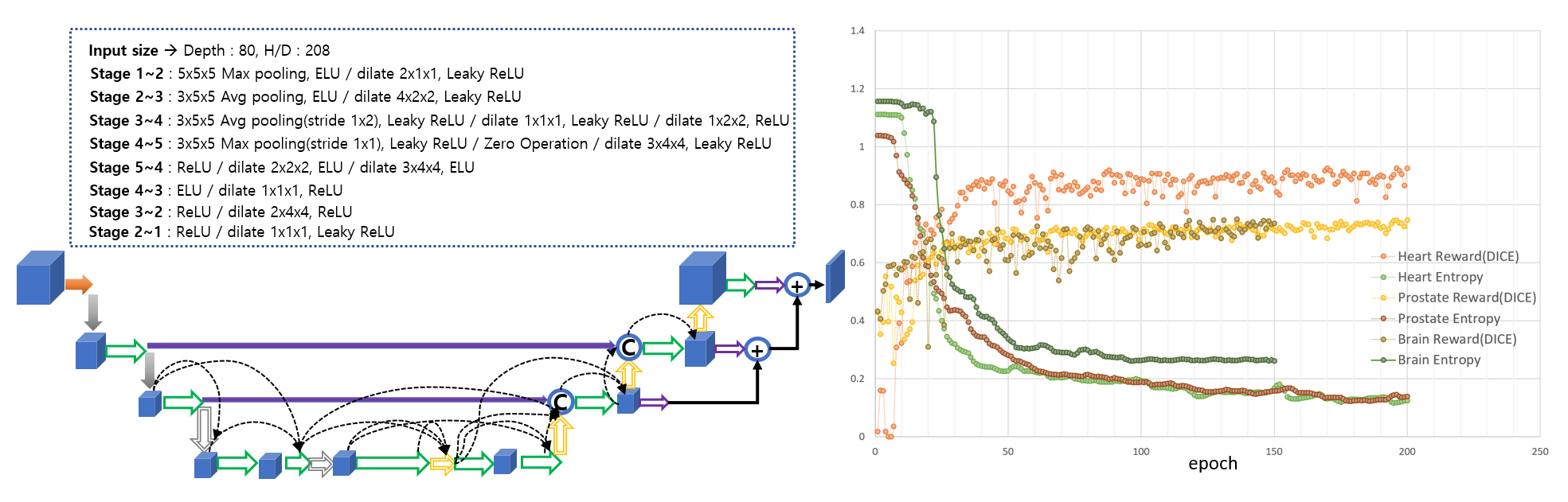}  %0.28
	\caption{ \textbf{Left} : One of the optimal architectures chosen for the validation set of heart task. \textbf{Right} : The entropy of the controller and the reward of the selected architecture for each task.}
    \label{fig:searched_network}
\end{figure}

Figure \ref{fig:searched_network} shows the entropy and reward sequence of each task and the resulting optimal architecture obtained for the heart task for a particular validation set.
The uniformly decreasing entropy and steadily increasing rewards illustrate that the controller network was trained well and the result of the architecture is stable. In other words, the controller repeatedly suggests either the same or similar child networks producing consistent dice scores. In our experiments, the controller suggests the same network more than 80\% of the time regardless of the validation set.  

\setcounter{table}{1}
\begin{table}[]
\centering
\label{tab:DICE_result}
\caption{ Mean Dice score for Brain tumor, Heart, and Prostate 3D segmentation tasks. V.D.A, Ensemble, T.T.A, and P.P indicate whether Various Data Augmentation, Ensembling, Test Time Augmentation, and Post-Processing were used to obtain the final result. } %\textbf{RIL} 
\begin{adjustbox}{max width=\textwidth} %\textwidth max width=0.95\textwidth
\begin{tabular}{|c|c|c|c|c|c|}
\Xhline{4\arrayrulewidth}
\hline
 & 3D U-ResNet\cite{Sungwoong_SCNAS} & SCNAS\cite{Sungwoong_SCNAS} & \begin{tabular}[c]{@{}c@{}}SCNAS\\ ( transfer )\end{tabular} & nnUnet\cite{isensee2018nnu} & \begin{tabular}[c]{@{}c@{}}RONASMIS \\ ( non fine-tunning )\end{tabular}  \\ \hline
Brain Tumor & 71.61 & 72.04 & - & 74.00 & \textbf{74.14}    \\ \hline
Heart & 89.60 & 89.99 & 90.47 & 92.70 &  \textbf{92.72}    \\ \hline
Prostate & 63.77 & 65.30 & 67.92 & 74.54 &  \textbf{75.71}  \\ \hline
\Xhline{2\arrayrulewidth}
\begin{tabular}[c]{@{}c@{}}V.D.A, Ensemble, \\T.T.A, P.P.\end{tabular} & No & No & No & Yes &  \textbf{No} \\ \hline
Training GPU & Tesla V100 & Tesla V100 & Tesla V100 & - &  \textbf{One RTX 2080Ti} \\ \hline
Inference of network & Overlapped patch-wise & Overlapped patch-wise & Overlapped patch-wise & \begin{tabular}[c]{@{}c@{}}Weighted \\overlapped patch-wise\end{tabular} &  \textbf{One-shot} \\
\Xhline{4\arrayrulewidth}
\end{tabular}
\end{adjustbox}
\end{table}

\section{Conclusion}\label{conclusion}
By configuring an efficient search space using macro search and utilizing parameter sharing for training a controller, we were able to apply NAS to 3D medical imaging segmentation tasks where previously developed NAS methods were difficult to apply. The proposed resource-optimized NAS framework outperforms state-of-the-art results obtained by manual design in the 3D medical image segmentation challenge. Furthermore, our proposed method is more meaningful in that it achieves excellent performance without
using various data augmentation, ensembling, T.T.A, and post-processing.

\section{Acknowledgement}\label{Acknowledgement}
This research was supported by a grant of the Korea Health Technology R\&D Project(grant number : HI18C0673) through the Korea Health Industry Development Institute (KHIDI), funded by the Ministry of Health \& Welfare, Republic of Korea and Industrial Strategic technology development program (grant number : 10072064) funded by the Ministry of Trade Industry and Energy, Republic of Korea.

\bibliographystyle{splncs03}

\bibliography{RONASMIS}

\begin{thebibliography}{10}
\providecommand{\url}[1]{\texttt{#1}}
\providecommand{\urlprefix}{URL }

\bibitem{bender2018understanding}
Bender, G., Kindermans, P.J., Zoph, B., Vasudevan, V., Le, Q.: Understanding
  and simplifying one-shot architecture search. In: International Conference on
  Machine Learning. pp. 549--558 (2018)

\bibitem{chen2018encoder}
Chen, L.C., Zhu, Y., Papandreou, G., Schroff, F., Adam, H.: Encoder-decoder
  with atrous separable convolution for semantic image segmentation. In:
  Proceedings of the European Conference on Computer Vision (ECCV). pp.
  801--818 (2018)

\bibitem{dou20163d}
Dou, Q., Chen, H., Jin, Y., Yu, L., Qin, J., Heng, P.A.: 3d deeply supervised
  network for automatic liver segmentation from ct volumes. In: International
  Conference on Medical Image Computing and Computer-Assisted Intervention. pp.
  149--157. Springer (2016)

\bibitem{drozdzal2016importance}
Drozdzal, M., Vorontsov, E., Chartrand, G., Kadoury, S., Pal, C.: The
  importance of skip connections in biomedical image segmentation. In: Deep
  Learning and Data Labeling for Medical Applications, pp. 179--187. Springer
  (2016)

\bibitem{isensee2018nnu}
Isensee, F., Petersen, J., Klein, A., Zimmerer, D., Jaeger, P.F., Kohl, S.,
  Wasserthal, J., Koehler, G., Norajitra, T., Wirkert, S., et~al.: nnu-net:
  Self-adapting framework for u-net-based medical image segmentation. arXiv
  preprint arXiv:1809.10486  (2018)

\bibitem{liu2018darts}
Liu, H., Simonyan, K., Yang, Y.: Darts: Differentiable architecture search.
  arXiv preprint arXiv:1806.09055  (2018)

\bibitem{mortazi2018automatically}
Mortazi, A., Bagci, U.: Automatically designing cnn architectures for medical
  image segmentation. In: International Workshop on Machine Learning in Medical
  Imaging. pp. 98--106. Springer (2018)

\bibitem{pham2018efficient}
Pham, H., Guan, M.Y., Zoph, B., Le, Q.V., Dean, J.: Efficient neural
  architecture search via parameter sharing. arXiv preprint arXiv:1802.03268
  (2018)

\bibitem{ronneberger2015u}
Ronneberger, O., Fischer, P., Brox, T.: U-net: Convolutional networks for
  biomedical image segmentation. In: International Conference on Medical image
  computing and computer-assisted intervention. pp. 234--241. Springer (2015)

\bibitem{Sungwoong_SCNAS}
S, W~Kim, I.K.e.a.: Scalable neural architecture search for 3d medical image.
  https://openreview.net/pdf?id=S1lhkdKkeV  (2019)

\bibitem{shah2018ms}
Shah, M.P., Merchant, S., Awate, S.P.: Ms-net: Mixed-supervision
  fully-convolutional networks for full-resolution segmentation. In:
  International Conference on Medical Image Computing and Computer-Assisted
  Intervention. pp. 379--387. Springer (2018)

\bibitem{wang2017automatic}
Wang, G., Li, W., Ourselin, S., Vercauteren, T.: Automatic brain tumor
  segmentation using cascaded anisotropic convolutional neural networks. In:
  International MICCAI Brainlesion Workshop. pp. 178--190. Springer (2017)

\end{thebibliography}
%\begin{thebibliography}{11}
%\end{thebibliography}

\end{document}